# Integrating molecular models into CryoEM heterogeneity analysis using scalable high-resolution deep Gaussian mixture models

Muyuan Chen1, Bogdan Toader2, Roy Lederman2

1 Division of CryoEM and Bioimaging, SSRL, SLAC National Accelerator Laboratory, Stanford University, Menlo Park, CA, USA
2 Department of Statistics and Data Science, Yale University, New Haven, CT, USA

## Abstract

Resolving the structural variability of proteins is often key to understanding the structure-function relationship of those macromolecular machines. Single particle analysis using Cryogenic electron microscopy (CryoEM), combined with machine learning algorithms, provides a way to reveal the dynamics within the protein system from noisy micrographs. Here, we introduce an improved computational method that uses Gaussian mixture models for protein structure representation and deep neural networks for conformation space embedding. By integrating information from molecular models into the heterogeneity analysis, we can resolve complex protein conformational changes at near atomic resolution and present the results in a more interpretable form.

## Graphical abstract

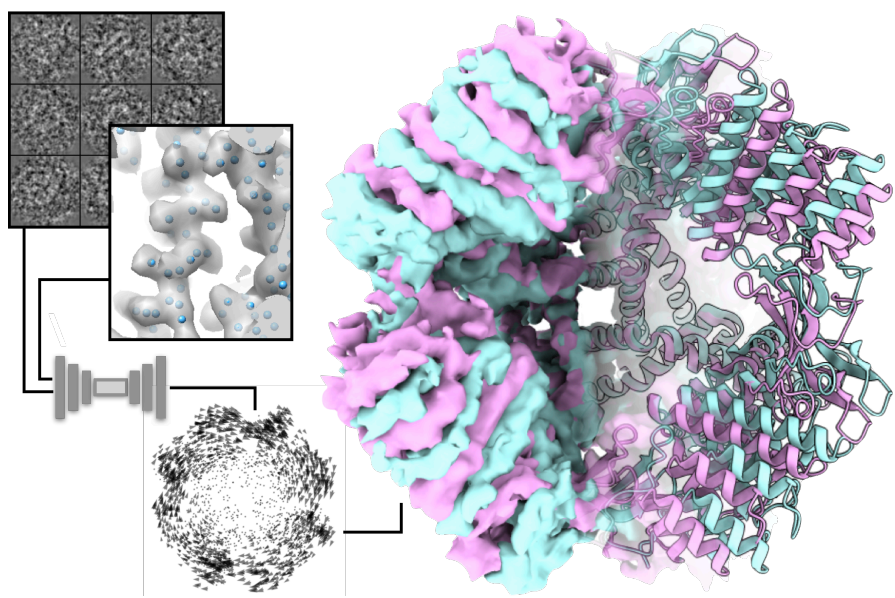

# Introduction

Inside cells, proteins and other macromolecules perform various tasks through dynamic interactions among their own components or with other molecules. Studies of the structural flexibility of those macromolecules are often key to understanding how they accomplish their functions. Cryogenic electron microscopy (CryoEM) takes snapshots of macromolecules frozen in vitrified ice, providing direct information about the compositional/conformational states of individual protein particles. While the structural heterogeneity is often a limiting factor for the high resolution structure determination using CryoEM, with the help of advanced computational methods, it also presents the opportunity to directly visualize the dynamic process of protein assembly and conformational changes, leading to insights on the functioning mechanism of the macromolecules.

Over the past few years, multiple methods have been developed that address the structural heterogeneity in CryoEM. Both traditional statistical inference approaches and deep learning based methods have been implemented to tackle this problem, and have achieved success in analyzing various real CryoEM datasets[1–4] (a detailed review of them can also be found in this special issue). Among those methods, we focused on the development of computational protocols that represent the structure of proteins as a Gaussian mixture model (GMM) and resolve the compositional and conformational heterogeneity of the proteins using deep neural networks (DNN)[5]. By representing the protein density map as the sum of many Gaussian functions, the method greatly reduces the computational complexity of the problem and makes it possible to capture the structural variability of highly dynamic systems. For near atomic resolution structures, the GMM representation often requires less than 1% of the resource compared to the voxel based representation (Fig.1A). The Gaussian representation also provides a way for the researchers to focus the heterogeneity analysis on individual domains of interest, making it easier to interpret complex conformational changes within the system.

Gaussian models provide a natural way to bridge CryoEM density maps and molecular models. In fact, to measure the similarity between the CryoEM map and the corresponding atomic model, one typical way is to generate a density map by placing a Gaussian function for each atom in the model and computing the FSC between the simulated map and the CryoEM structure[6]. Additionally, Gaussian falloff at the location of each atom is also used to measure the quality of the CryoEM structure[7,8]. In this work, we further exploit the connection between the GMM representation of protein structures and the molecular models and develop new algorithms that guide the heterogeneity analysis of CryoEM datasets using information from the models. Prior information from models helps overcome low SNR in individual particles at high resolution and provides extra constraints for the analysis of complex systems.

To make the model guided heterogeneity analysis possible, we start by developing a memory efficient implementation of the GMM, which can now represent protein structures and dynamics at near atomic resolution. With the new implementation, we show three distinct ways to integrate information from molecular models into heterogeneity analysis. First, a hierarchical GMM is used to model the large scale global movement within the target protein at near atomic resolution. Second, an alternative DNN architecture is adopted to model the localized domain motion as rigid body movement. Finally, we introduce an approach that regularizes heterogeneity analysis at high resolution using bond constraints from the corresponding molecular models. In addition, we also present a way to combine multiple modes of heterogeneity analysis methods for the same system, focusing on different parts of the protein and different resolution ranges, to provide a more comprehensive view of the dynamics of macromolecules.

## Methods

Following the previous implementation of the method, a pair of DNNs is used to analyze the structural variability from a CryoEM single particle dataset. The first DNN, the encoder, takes the information from individual particles and maps them onto a small latent space where each point represents the conformation of one particle. To ensure the continuity of the latent space, a small random variable following Gaussian distribution is added to the output from the encoder, mimicking the behavior of a variational autoencoder, except that the width of distribution is not trainable. The second DNN, the decoder, takes the coordinates from the conformational space, and outputs a set of parameters that define a GMM, including the 3D coordinates, amplitude and width for each Gaussian function. Then, a projection image is generated from the GMM at the same orientation of the input particle, and the similarity between the GMM projection and the particle image is used as the loss function for the training of the encoder-decoder pair. Using the autoencoder architecture, the DNNs can be trained to learn the dynamics within the protein system from a set of particles, whose orientations are predetermined from a single model refinement, without human supervision.

### Memory efficient implementation of GMM

One difficulty in the previous implementation is the limitation of GPU memory. While the GMM representation itself only requires minimal memory usage, a large block of memory still needs to be allocated when generating 2D images from the GMM. For N Gaussians and image dimension of M x M pixels, the previous implementation consisted of computing the discretized projection of each Gaussian and pixel-wise adding them to obtain the final projection image, which requires the storage of N x M x M pixel values. A more efficient way to implement the projection of N Gaussians on a M x M grid is to project the N Gaussians on the x and y axes separately, which requires the storage of two N x M matrices, that are then multiplied to obtain the 2D projected image. This approach has been used in the context of CryoEM for obtaining projection images of a GMM in the real domain[9], while in this work we perform the projection in the Fourier domain.

### Building GMMs from molecular models

In the previous work, the GMM was directly built from the neutral structure by training the decoder from scratch using the projection images. While this approach has stable performance for low to intermediate resolution structures, the convergence of the training process can be slow when the GMM involves thousands of Gaussian functions, and the structure is determined at near atomic resolution. To speed up the GMM generation process and enable the utilization of molecular model information, we re-designed the process so the decoder can be initialized from existing molecular models. Either full atom models of the protein or pseudoatom models seeded from the density map can be used to initialize the training of the decoder. In some situations, particularly when the CryoEM structure contains unmodeled densities, the two approaches need to be combined to produce the initial GMM. For example, to build the GMM of membrane proteins embedded in nanodiscs, in addition to the molecular model of the protein, extra Gaussian functions need to be seeded on the lipid density to fully represent the structure.

To build the GMM from molecular models, we first fit the model to the density map of the neutral structure in visualization software such as UCSF Chimera[10]. When unmodeled densities are present, we subtract the density around the existing atoms from the neutral structure, then seed pseudoatoms on the unmodeled density map using a K-means algorithm on the coordinates of remaining voxels in 3D real space[11]. Then, to build the decoder, we initialize its weights with random small values and train it using gradient descent so that the coordinates of the Gaussian functions from its outputs match the coordinates of the atoms from the input model. Meanwhile, the amplitude and width of Gaussian functions are left constant during this initial training. Since there is only one set of noise free target outputs, the initial training converges rapidly, normally within a minute. Finally, starting from the pre-trained decoder, we perform another round of iterative training to minimize the difference between the projection images generated from the output GMM of the decoder and the projections from the neutral 3D structure (Fig.1B). The Gaussian coordinates are locally optimized to fit the density map, and the amplitude and width of the Gaussian functions are adjusted to match the occupancy and local resolvability of the corresponding protein domains. In addition, bonds between the atoms from the molecular models can also be considered during the training process as regularization factors, so that the resulting GMM satisfies the biochemical constraints. Detailed implementation of the bond constraints will be discussed in the following sections.

**Large scale global morphing with hierarchical GMMs**

In many situations, even when the structure of a protein is determined at near atomic resolution and atomic models can be built from the structure, global, large-scale movement can still be present within the system[12–14]. While most of the particles are at the conformations close to the neutral structure, making high resolution structure determination possible, a small fraction of particles still undergoes significant conformational changes that can be observed at low resolution. The large-scale movement at the domain level will further drive the movement of secondary structure elements (SSEs) within the domain, as well as the motion of high resolution features such as the loops connecting the SSEs and the sidechain of each residue. To have a comprehensive view of the structural heterogeneity of a protein, it is critical to study the system in a hierarchical way. However, due to the complexity of protein movement and the nature of DNNs, it is challenging to train a model that describes the conformational change of the protein at all resolution ranges in a single step approach. To address this problem, we designed a hierarchical set of GMMs that are embedded in the DNN architecture, which connects the large-scale morphing of the protein to the corresponding movement of high resolution features.

To model the multi-scale movement of the protein complex hierarchically, we start by building two GMMs from the neutral structure of the protein: a large GMM with thousands of Gaussian functions that match the protein density map at near atomic resolution, and a small one with only a few (<100) Gaussian functions that fit the density map at a low resolution where the large scale movement can still be observed. Then, we build an MxN transition matrix based on the distance matrix between the Gaussian coordinates of the two GMMs, where M and N are the numbers of Gaussian functions in the small and larger Gaussian models. The weights in the matrix are initialized as exp(-50*D^2), where D is the pairwise distance between the two GMMs. As such, by multiplying a set of vectors that describe the movement of the small GMM with this matrix, each of the vectors can drive the movement of Gaussian functions in a local region from the large GMM, enabling the coordinated movement of the two GMMs (Fig2.A-D).

In the first round of heterogeneity analysis, the initial decoder output is an Mx5 matrix of small random values. The output is then multiplied by the MxN transition matrix, and the resulting Nx5 matrix is added to the large GMM built from the neutral structure. The initial training process compares the projection images from the large GMM and the particle images, using a Fourier ring correlation (FRC) loss function that cutoff at low resolution. At this stage, since the weights in the transition matrix are constant, and we only focus on the low resolution signal, it is relatively easy for the encoder-decoder pair to converge to an optimal solution. After the first round of training converges, we move on to include the high resolution signal in the heterogeneity analysis. In the second round of training, we convert the transition matrix into a trainable dense layer of the decoder and append it to the existing layers of the DNN. As a result, the decoder now directly outputs an Nx5 matrix that can alter the coordinates and amplitude of the large GMM corresponding to the low resolution conformational change it learned from the first round of training. Finally, we train the encoder-decoder pair again, this time including the weights in the transition matrix as trainable variables and using the FRC of the full resolution range as the loss function. The second round of training refines the low resolution movement the DNNs captured in the first round, and extracts the movement of high resolution features corresponding to the large scale conformational change.

**Modeling localized domain movement using rigid body motion**

One set of problems often encountered in CryoEM studies is proteins with rigid cores but a few highly flexible domains[15–17]. The core parts of the protein remain consistent among most particles, thus the structure can be determined at high resolution, while the flexible domains are smeared out in the averaged structures. To resolve the large-scale, localized movement of the flexible domain, we model the movement of one domain with respect to the rest parts of the protein as a rigid body motion. By assuming the features within the target domain rotate and translate as a whole, the problem of heterogeneity analysis is greatly simplified, and it becomes easier to model highly nonlinear and long-range conformational changes. The properties of Gaussian functions make it convenient to represent the system in either real or Fourier space, providing a good way to focus the heterogeneity analysis on any specific domain of the protein. The rigid body transform of a target domain can be simply represented by shifting the coordinates of Gaussian functions corresponding to that domain while keeping the location of other Gaussian functions unchanged. Since the full GMM is used to generate the projection images, this will not produce any seam line between the target domain and the rest parts of the protein, or any mask induced artifacts in the Fourier space.

The first step for rigid body motion based heterogeneity analysis is to define the domain by selecting a group of Gaussian functions in the model. This can be done by providing a volumetric mask to select all Gaussian functions whose centers fall under the mask, or by specifying indices of atoms in the molecular model. When there are existing models of the target domain, they can also be used in building the initial neutral state GMM, as well as defining the boundary of the domain. This can be particularly helpful when large scale movement is present in the system, since the smearing in the neutral model can alter the size and shape of the target domain in the neutral structure density map.

To model the domain movement as rigid body motion, instead of parameters for a GMM, we redesign the decoder, so it maps each point in the latent conformation space to a 1x6 vector, representing the rotation-translation parameters for the target domain for one single particle. A 3D rotation matrix can be generated from the vector, and multiplying this matrix with the coordinates of Gaussian functions within the domain will transform the target domain according to the parameters from the decoder output. During the heterogeneity analysis, the transformed target domain, combined with the neutral state GMM of the rest parts of the protein, is used to generate the projection images. Finally, training loss is computed by comparing the projection images with the particles, and the encoder-decoder pair is optimized to capture the trajectory of the rigid body movement of the target domain (Fig.2E-G).

**High resolution heterogeneity analysis with bond constraints**

In addition to the coordinates of atoms from the molecular model, the connectivity between the atoms can also be used as prior information to guide GMM-DNN based heterogeneity analysis. The usage of bond constraints ensures that the decoder prefers GMMs at biochemical viable conformations, making it easier to extract biologically meaningful information from the noisy particle images. Here we focus on the C-alpha backbone models of the target protein, and use the distance between neighboring C-alpha atoms, as well as the angle between each C-alpha-C-alpha pseudobond to guide the heterogeneity analysis at high resolution.

To utilize the bond information, we first calculate the distance between every adjacent C-alpha atom and the angle between each C-alpha pseudobond from the existing backbone model at the neutral state. When an existing full atom model is present, the backbone model can be parsed from the atomic model. Otherwise, the backbone model can also be built automatically from the CryoEM density map using various software tools[18,19]. During the training process, both when refining the neutral state GMM to fit the averaged structure, and during heterogeneity analysis using the particles, we calculate the same C-alpha distances and angles each time a GMM is generated by the decoder. The distances and angles are compared to their corresponding values from the neutral backbone model, and a small regularization term is added to the training loss based on the average difference between the neutral model and the GMM produced by the decoder at any conformational state. The regularization term forces the output GMM to match the conformation of individual particles without significantly altering the geometry of the backbone model locally. For example, individual alpha-helices can tilt as part of a conformational change, but the C-alpha atoms along the helix will still generally keep an alpha-helix formation during such motion.

**Coordinating structural heterogeneity analysis of multiple modes**

In a complex protein system, very often multiple modes of conformational change, such as the movement of two separate domains, exist simultaneously. To have a thorough understanding of the structural dynamics of the protein system, it is necessary to not only study each movement mode separately, but also investigate the correlation between the different modes of structural variability. The masking functionality in the GMM-DNN based analysis already provides a convenient way to study the structural heterogeneity of different parts of the protein independently, and here we also present a protocol that combines multiple DNNs to study the coordination of the different modes of movement within the protein.

Theoretically, it is possible to simply focus the heterogeneity analysis on different parts of the protein independently, and study the correlation between the movement modes by analyzing the particle distribution in multiple conformational spaces produced by the different encoders using statistical methods. However, given the complexity of the protein conformational changes, the noise in particle images, as well as the nonlinearity in the embedding methods, direct correlational analysis of the encoder outputs may not yield meaningful information. To address this issue, we study the coordinated movement by combining the encoders while keeping the separate decoders. That is, we use only one encoder that maps the information of the particles to a conformational space, combined with multiple decoders that take the same conformation input from the encoder, but map it to the conformational changes at different parts of the protein. Since the decoders are entirely independent, different architectures, including the new functions listed above, can be used to address different types of heterogeneity in the target domain. For example, we can have one decoder focusing on the large scale rigid body movement of a specific domain, and another one that uses backbone model constraints for the more subtle conformational change in a more rigid part of the protein. The output of each decoder would be the offset of GMM parameters with respect to the neutral model, so for each particle, the results from multiple decoders can simply be summed together to generate the GMM representing the conformation of that particle. In practice, to facilitate the convergence of the heterogeneity analysis, we start by optimizing the encoder and only one decoder to model the conformational change within one domain, normally the one with the largest scale of movement. Then, in the second round of training, we can include other decoders to extract the more subtle structural changes in other parts of the protein that correlate with the motion in the first target domain.

## Results

### TRPV1 (EMPIAR-10059)

The first example we used to demonstrate the improvement of the heterogeneity analysis method is a single particle dataset of TRPV1 embedded in nanodisc[12]. From this public dataset, the structure of TRPV1 can be determined at 3Å resolution with C4 symmetry from ~200,000 particles. While the transmembrane domains of the protein show apparent high resolution features, the ankyrin repeats (AR) domain outside the lipid nanodisc was not as well resolved. So, in this example, we used the hierarchical GMM approach and targeted the global, large-scale movement of the AR domain.

First, a GMM with 8000 Gaussian functions was constructed to match the neutral state density map at 3Å resolution (Fig.3A-B). This was done by first seeding pseudoatoms from the density map, then converting it to a GMM and refining it against the projection images of the neutral structure. Meanwhile, a small GMM with only 32 Gaussian functions was also built using the same method that represents the same neutral structure at 20Å resolution. A hierarchical model was constructed using these GMMs, and used for the heterogeneity analysis of the particles. At the beginning of the analysis, we expanded the symmetry of the structure to C1 by duplicating each particle four times at the four symmetrically equivalent orientations. The first round of analysis only optimized the parameters of the small GMM, and used particle-projection FRC cutoff at 15Å as the loss function. After the training converges, the second round of heterogeneity analysis was performed on unbinned particles to refine the movement of the fine features. Along one of the eigenvectors in the resulting conformation space, the helices in the AR domain underwent a global rotation of 6 degrees, while the helices in the transmembrane domain remained static. Other interesting conformational changes can also be observed along other vectors in the conformation space, including the movement of AR domains on the opposite sites toward the center of the complex, as well as the global tilting of the domains from all four asymmetrical units (Fig.3C).

After the GMM based heterogeneity analysis, we converted the movement trajectory of the GMM to the movement of the molecular model to better visualize the different modes of conformational changes in 3D. This was done by simply treating the molecular model of the protein as another layer of the hierarchical GMM. Starting from an existing atomic model of TRPV1 fitted into the neutral state density map, we constructed a transition matrix between the large GMM and the coordinates from the atomic model, the same as the transition matrix between the small and large GMM. By multiplying the GMM offsets produced by the decoder with the transition matrix and adding the offset to the neutral state atomic model, we can morph the molecular model to produce structures at any state from the conformation landscape. The models generated using this method fitted well into the density maps reconstructed from particles around the same conformation, and provided a quantitative description of the conformational changes within the system (Fig3.D-E).

The peak usage of GPU memory during the heterogeneity analysis of the TRPV1 dataset was 9.8GB, within the capacity of modern consumer GPUs, indicating that the proposed method is capable of dealing with large datasets and high resolution structures. In addition, to reduce the usage of CPU memory during the analysis, instead of loading all particles into the memory and feeding them to the DNN, we broke the dataset into chunks of 10,000 particles, and process them sequentially. For each chunk of particles, the program loaded the DNNs trained from the previous rounds, trained for ten iterations on current particles, and saved the trained DNNs to the hard drive. Then, both CPU and GPU memory usage was cleared, and the program moved on to the next chunk of particles. For each chunk of particles, a subset of particles (10% by default) was separated to form a test set, in order to monitor the convergence of the training process. Finally, after the training converges, the program went through all particles one more round to compute the conformation of each particle using the saved encoder.

**GLP-1 receptor (EMPIAR-10346)**

In the second example, we used a public dataset of the GLP-1 receptor to show the new functionality of rigid body movement and model guided heterogeneity analysis[20]. While an averaged structure can be determined at 3.8Å resolution from the ~500,000 particles through a homogeneous refinement, the extracellular domain (ECD) nearly vanished in the high resolution structure, and some helices in the transmembrane domain were not as well resolved. Here, we analyzed the large scale movement of the ECD using the rigid body analysis, and modeled the coordinated conformational change of the transmembrane helices using the bond constraints from the molecular model.

To build the GMM, we first seeded Gaussian coordinates at C-alpha positions according to the full atom model (PDB:6ORV) deposited with the dataset. Then, the neutral state density map was filtered to 10Å so the densities of lipid nanodisc and the ECD show up, and the remaining Gaussian coordinates were placed on the unmodeled densities. The two GMMs, which include 2100 Gaussian functions, were combined and refined against the neutral state structure targeting 4Å resolution. The C-alpha bond length and angle constraints, which were derived from the molecular model, were used to regularize the refinement of neutral GMM (Fig.4A-B). Two decoders were used to model the conformational change within the system, one targeted the rigid body motion of the ECD, and the other focused on the movement of the transmembrane helices. The heterogeneity analysis started from training the encoder together with the first decoder, which outputs the rotation/translation parameters for the ECD with respect to the rest of the protein, targeting 15Å resolution to extract the large-scale domain motion from the particles. After the first round of training converged, we included the second decoder and start another round of training. During training, the second decoder was only allowed to adjust the position and amplitude of the Gaussian functions corresponding to the transmembrane helices, and the C-alpha bond constraints were included as a regularizer.

Along the motion trajectory represented by the first eigenvector from the conformational space, the ECD was tilting up and down with respect to the membrane plane, while also swinging horizontally (Fig.4C-D). The overall rotation of the ECD along the motion trajectory was around 30 degrees. Correlating to the tilting of the EDC, movement of the transmembrane helices can also be observed. The extracellular end of transmembrane helix 1 (TM1), which directly connects the ECD, shifted by as much as 8Å from the first to the last frame of the trajectory. Additionally, large movement at the extracellular end of TM3 and TM6 can also be observed along the trajectory (Fig.4E).

**Discussion**

One of the major difficulties in single particle heterogeneity analysis is the low signal-to-noise ratio (SNR) in individual particle images. The problem becomes even more challenging when we look at the structural variability of proteins at high resolution, as more nonlinearity is required to model the conformational changes of fine protein features, and the SNR decays even further at high spatial frequencies. To facilitate the structural variability study of macromolecules at high resolution, the latest development in the GMM-DNN based heterogeneity analysis methods focuses on reducing the complexity of the problem using prior information from molecular models. By focusing on the large-scale morphing of a high resolution model, or the rigid body motion of a specific domain, we reduce the scale of the solution space to facilitate the convergence of the DNNs. The bond information from the molecular models provides extra constraints for the analysis, which is particularly useful in extracting biochemically meaningful protein conformational changes, instead of the random variability in particle images caused by noise.

The sheer complexity of the dynamics within the protein complexes presents another difficulty for the heterogeneity analysis. Often a single domain of a protein can exhibits continuous conformational changes with multiple degrees of freedom, and multiple domains of the same protein can move independently, or coordinate to perform a single function. The high complexity of the system makes it harder to train the DNNs to converge, and even when the training converges, it can be challenging for human researchers to decipher the latent space of the DNNs and make meaningful conclusions about the protein system. The masking capability of the GMM method makes it possible to analyze a highly dynamic system using a divide-and-conquer approach. Building one encoder-decoder pair focusing on one individual domain makes it easier for the training process to converge without intensive hyperparameter tuning, and also more convenient for a human to inspect and rationalize the conformational changes learned by the DNNs. Furthermore, the multi-decoder approach introduced in this work makes it possible to investigate the correlation between the movement of different domains, and obtain a more comprehensive view of the dynamics of the macromolecular system.

Building GMMs according to the molecular models also makes it possible to convert a GMM at any conformational state back to the corresponding molecular model. This provides a convenient way to directly visualize the conformational changes learned by the DNNs. Since the DNNs consolidate the information from all particles as well as the biochemical constraints to produce the conformation space, the model generated from any individual point from the latent space can contain more information than the averaged structure reconstructed from nearby particles in the latent space. This feature is particularly useful when multiple degrees of freedom are present in the system so that the particle count in each individual state is low, or when we are interested in a transient state along a path of conformational change.

The development of various new functionalities for the GMM based heterogeneity analysis presented in this work also leads to a flexible software platform, which is built to test the performance of different modules of the protocol. To speed up training, small sample datasets with clear structural variabilities are constructed to test the alternative algorithms. The platform is implemented in a modular form using Jupyter notebook, and each step can be easily replaced to test the alternative methods. The platform, along with the documentation and sample datasets, will be distributed in near future, so other method developers can implement their new heterogeneity analysis algorithms under this framework. The software tool is distributed with EMAN2[21], and corresponding documentation can be found through eman2.org.

**Acknowledgement**

This work was supported by NIH grant R01GM080139, R21MH125285, R01GM136780, as well as FA9550-21-1-0317 from AFOSR. Computational resources from SLAC Shared Scientific Data Facility are used for this work.

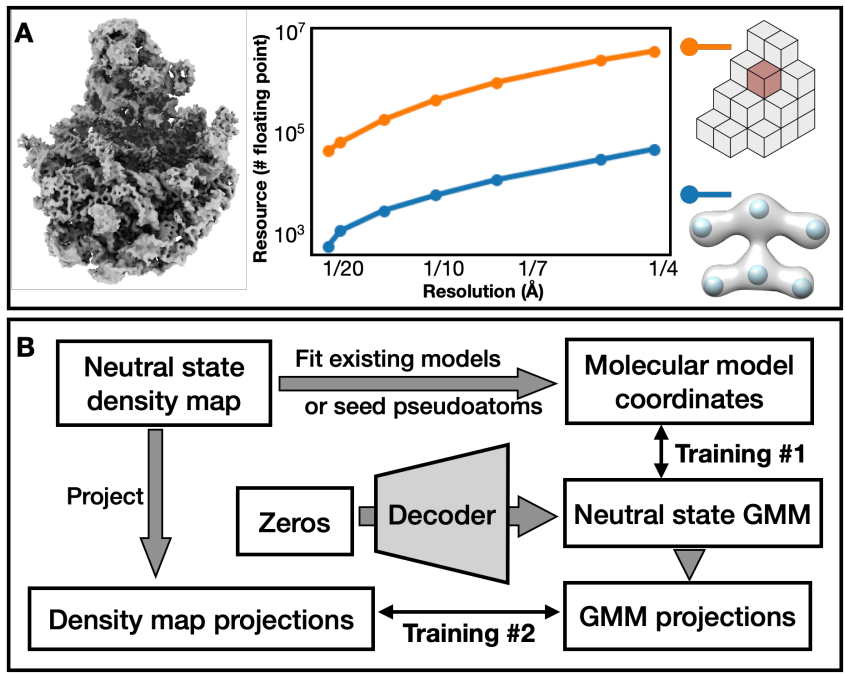

**Fig.1. Building Gaussian models from neutral state density maps. (A)** Resource usage comparison between the voxel based representation and the GMM based one. Note the y axis in the plot is in log scale. **(B)** Workflow for building a neutral state GMM from the corresponding density map. The first round of training matches the output GMM from the decoder to the coordinates of the molecular model, and the second round of training maximizes the similarity between the projections of the density map and the projections of the GMM.

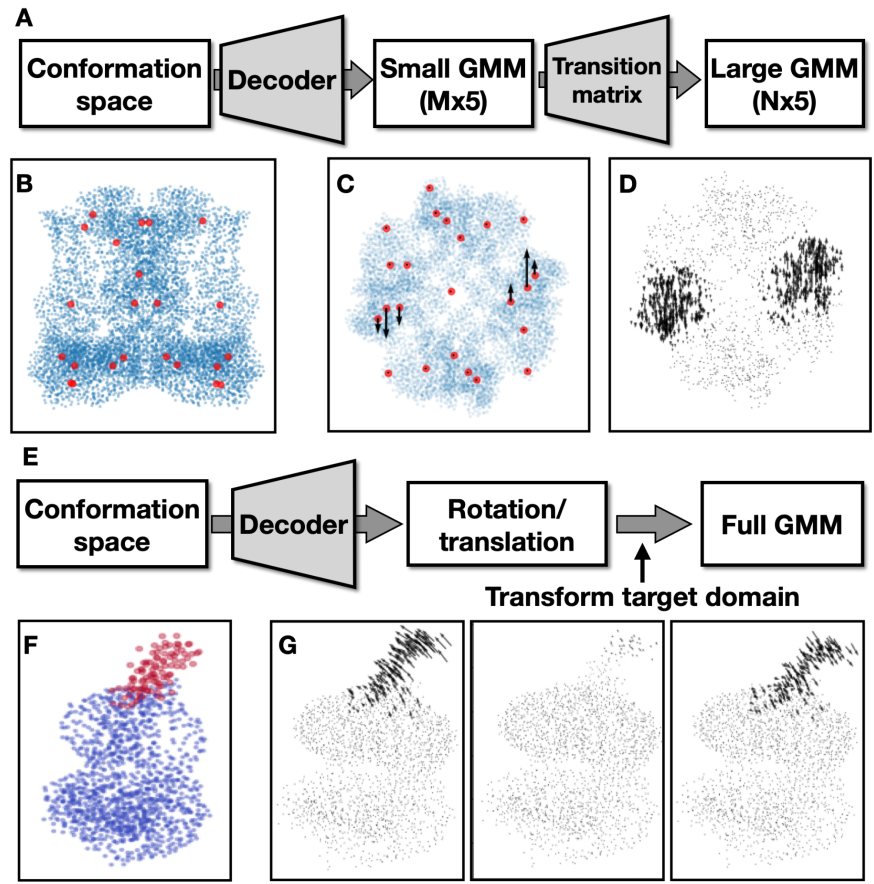

**Fig.2. Modeling large scale movement. (A)** Diagram hierarchical GMMs for global morphing. In the first round of training, target low resolution FRC and train only the decoder. In the second round of training, train weights in both the decoder and the transition matrix and target high resolution FRC. **(B)** Coordinates from the large GMM (blue) overlayed with the small GMM (red). **(C)** Example movement trajectory of the small GMM. **(D)** Quiver plot showing the morphing of large GMM driven by the movement in C. **(E)** Diagram of GMM-DNN based rigid body movement analysis. **(F)** Neutral GMM, with the target domain highlighted in red. **(G)** Learned eigenvectors of the rigid body movement of the target domain.

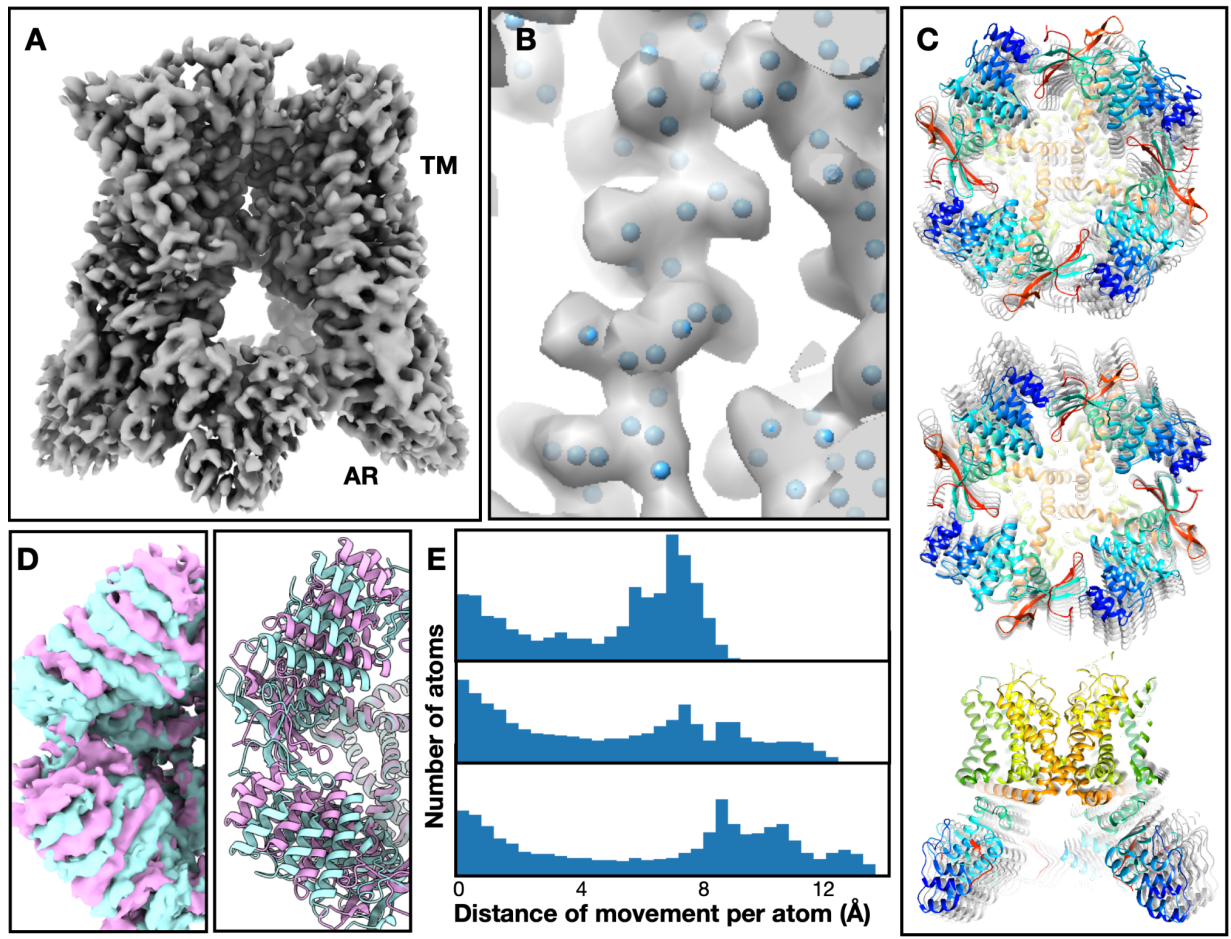

**Fig.3. Heterogeneity analysis results from the TRPV1 dataset. (A)** Neutral state structure. **(B)** Close-up view of a transmembrane helix from the neutral state GMM. Blue dots represent the coordinates of Gaussian functions used in the model. **(C)** Three different movement modes of the TRPV1 AR domain shown in morphed atomic models, including the rotation of AR domains around the symmetry axis, movement of opposite AR domains toward the center, and global tilting of all AR domains. **(D)** Comparison of the first movement mode visualized as reconstructed density maps and morphed atomic models. The two colors represent the first and last frames of the movement trajectory. **(E)** Quantifying the scale of different movement modes shown in C using histograms of per atom movement distance. The distance is measured from the morphed models at the first and last frames of the movement trajectory, covering 98% of the particles.

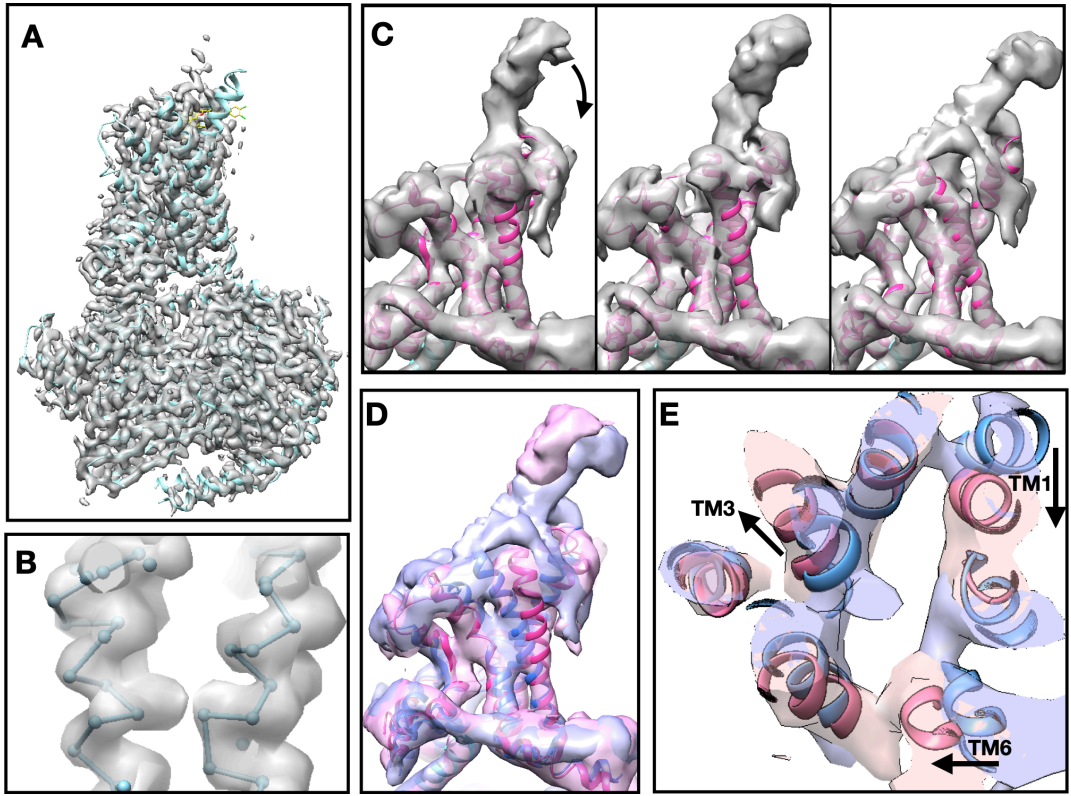

**Fig.4. Heterogeneity analysis results from the GLP-1 receptor dataset. (A)** Neutral state structure fitted with an existing molecular model (PDB:6ORV). **(B)** Close-up view of transmembrane helices from the neutral state GMM. Blue dots represent the coordinates of Gaussian functions and the bonds between them are used for regularization during DNN training. **(C)** Conformational change along the first eigenvector, with corresponding models fitted to the density maps. The maps are filtered to 5Å to emphasize the movement of helices. **(D)** Overlay of the first and last frame of C with corresponding molecular models. **(E)** Cross section of E, showing the movement of helices.